\documentclass[showpacs,twocolumn,prb, nofootinbib]{revtex4-1}
\usepackage{color}    

\usepackage[hmargin=0.69in, vmargin=0.75in]{geometry}              
\usepackage{graphicx}
\usepackage{amssymb}
\usepackage{amsmath}
\usepackage{epstopdf}
\usepackage{rotating}
\usepackage{natbib}
\usepackage[section]{placeins}
\usepackage{hyperref}

\usepackage[caption=false, position=t,singlelinecheck=off]{subfig}

\begin{document}

\title{
Theory and design of quantum light sources from quantum dots embedded in semiconductor-nanowire photonic crystal systems}
\author{Gerasimos Angelatos}
\email{g.angelatos@queensu.ca}
\author{Stephen Hughes}
\affiliation{Department of Physics, Engineering Physics and Astronomy, Queen's University, Kingston, Ontario, Canada K7L 3N6}
\begin{abstract}
We introduce  a new platform for realizing on-chip quantum electrodynamics
using  photonic-crystal waveguide structures comprised of periodic nanowire arrays with embedded semiconductor quantum dots to act as a quantum light sources.
 These  nanowire-based structures, which can now be fabricated with excellent precision, are found to produce
  waveguide Purcell factors exceeding 100 and on-chip $\beta$ factors up to 99\%.
We investigate the fundamental optical properties of  photonic crystal waveguides and finite-size structures, using both photonic band structure calculations and rigorous Green function computations which allow us to obtain the modal properties and the local density of photon states.
A comparison with slab-based photonic crystals is also made and we highlight a number of key advantages in the nanowire system, including the potential to reduce extrinsic scattering losses and produce high theoretical  Purcell factors and $\beta$ factors on-chip. We also demonstrate that these structures exhibit rich photonic Lamb shifts over broadband frequencies.

\end{abstract}
\maketitle

\section{Introduction}

A robust platform  for deterministically producing, manipulating, and transferring individual quanta ``on chip'' is a highly sought after commodity in quantum information science.  In particular, quantum cryptography \cite{Gisin2002} and optical quantum computation \cite{Biolatti2000,Ladd2010, Kimble2008} systems require a triggered single photon source \cite{Knill2001}.  Quantum dots (QDs),  semiconductor nanostructures which confine single excitons (electron-hole pairs) to a narrow spatial extent, can act as single photon emitters, with advantages such as large optical dipole moments, telecom-friendly emission wavelengths, and robustness in a solid state environment; however, they suffer from environmentally-induced decoherence,  excitation and collection issues \cite{Kiraz2004}.  One proposed solution to partially mitigate these effects is to implement QDs in a photonic crystal (PC) slab \cite{ Yao2009, Hennessy2007, BaHoang2012}, a periodic dielectric medium which controls the dispersion properties of light \cite{Yablonovitch1987, John1987, Johnson1999, Joannopoulos2011,  Krauss2008} in a planar slab geometry.  It has also been experimentally demonstrated that QDs can be embedded near PC cavity-field antinodes, dramatically increasing the spontaneous emission rate \cite{Hennessy2007, Yoshie2004} by modifying the local optical density of states (LDOS) experienced by the QD \cite{Lodahl2004}.  This allows one to realize ultrafast single photon sources, suppress decoherence, and design systems to study and exploit semiconductor cavity-quantum electrodynamics (QED) \cite{Carmichael1999, Yao2009, Bose2012}.

Photonic crystal waveguides\cite{Johnson2000} can also exploit this coupling enhancement due to a divergent LDOS at the waveguide mode edge, and have the added advantages of a  broader field enhancement bandwidth and directed photon emission, thus improving the ease of QD coupling and photon collection, respectively \cite{Rao2007theory,  Lecamp2007}. Consequently,   there has been active theoretical \cite{Hughes2004, Yao2009, Rao2007} and experimental \cite{Lund-Hansen2008, Dewhurst2010, Schwagmann2011, BaHoang2012} efforts towards the design of PC-waveguide-QD systems. In agreement with theoretical predictions \cite{Rao2007theory,Rao2007, Lecamp2007}, $\beta$ factors (the fraction of QD light emitted into a target mode) exceeding 90\% have been experimentally demonstrated \cite{Lund-Hansen2008}; however,  Purcell factors, i.e., the enhancement of the resonant QD spontaneous emission rate relative to a homogeneous medium,  above 3  have, to the best of our knowledge, yet to be demonstrated in ordered PC waveguide systems \cite{Dewhurst2010}. These modest Purcell factors are in part limited  from fabrication disorder inherent to the traditional PC platform, comprised of a periodic array of holes in a semiconductor slab.  Holes are formed in slabs using electron-beam lithography and etching, resulting in hole side-wall roughness and leading to significant 
scattering losses \cite{Hughes2005,Lund-Hansen2008}, particularly in the slow-light regime. Moreover, QDs in PC slabs are typically self-assembled through Stranski-Karatanow growth, limiting control over their position and emission frequency and resulting in poor coupling to PC waveguide modes  \cite{Yao2009,  BaHoang2012}.  This inherent randomness also prevents the design of structures containing multiple coupled QDs, a key requirement for many quantum information applications\cite{Biolatti2000, Ladd2010}.  

\begin{figure}[t]
\subfloat[]{\label{gaaswide}\includegraphics[width=0.23\textwidth]{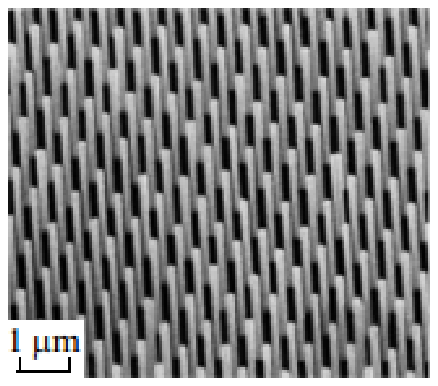}}\,
\subfloat[]{\label{realwgpic}\includegraphics[width=0.23\textwidth]{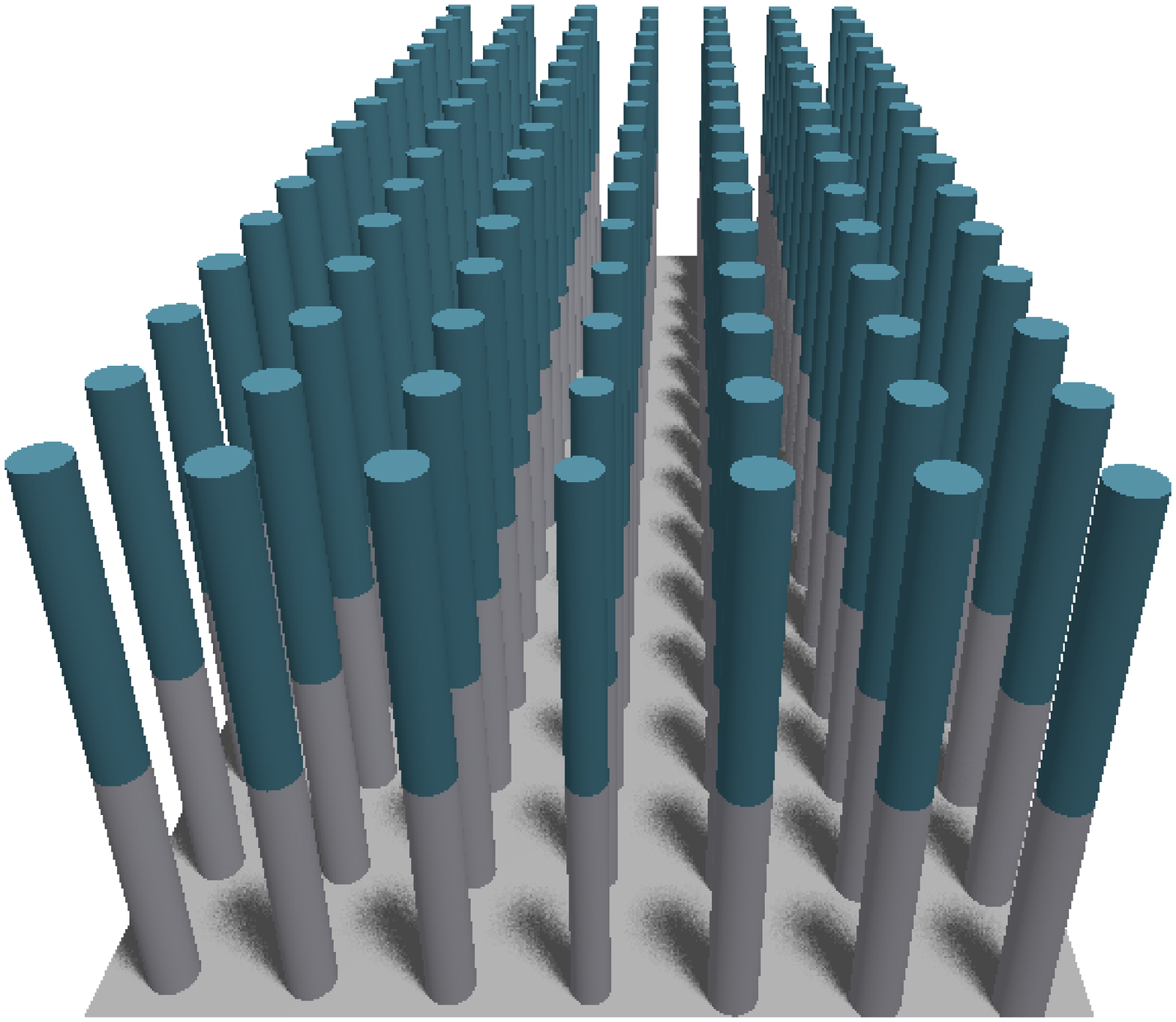}}

\vspace{-5pt}
\caption{{(a)   Nominally identical GaAs NWs grown in an organized structure from  Ref.~\onlinecite{Dubrovskii2009}. (b) Proposed NW waveguide, formed by reducing the radius of a single column of NWs in an array.  Guiding is done in the higher index (darker) upper portion of the NWs, while the lower index bottom separates the PC structure from the substrate below.}}
\label{pics}
\vspace{-15pt}
\end{figure}

The aforementioned slab design is, of course, not the only structure which can exploit PC physics; 
arrays of dielectric rods offer an alternative solid state system\cite{Johnson1999}.  Indeed, semiconductor nanorods or nanowires (NWs)  are being investigated for  a wide variety of photonics applications, such as single photon sources and detectors using embedded two-level systems \cite{Harmand2009, Babinec2010, Claudon2010}, optical cavities \cite{Birowosuto2014}, nanoscale lasers \cite{Duan2003}, and solar power collectors\cite{Dubrovskii2009,Czaban2009}.  Most NW work to date has focused on the electromagnetic properties of single NWs or disorganized ``forests''.  However,  recent techniques such as Au-assisted molecular beam epitaxy (MBE) have demonstrated the ability to fabricate large quantities of \textit{organized and identical} NWs \cite{Boulanger2011, Harmand2009, Dubrovskii2009, Makhonin2013}, an example of which can be seen in Fig.~\ref{pics}\subref{gaaswide}.  Due to the epitaxial growth process employed, single crystals are produced which will have atomically flat surfaces, suppressing scattering from surface roughness \cite{Dubrovskii2009}.  Furthermore, QDs can be embedded deterministically in NWs by tuning the growth conditions, allowing precise control of their size, position, and orientation \cite{Harmand2009, Tribu2008, Makhonin2013}.  Nanowire locations can be defined via electron beam lithography, allowing the  same  periodicity precision as seen in the traditional PC slab, but avoiding the associated structural damage.  We also note that techniques exist for adding QDs to the top of individual NWs post-process \cite{Pattantyus-Abraham2009}, enabling the design of systems coupling separated QDs at deterministic locations.

In this work we introduce the design of a new  platform  for semiconductor on-chip QED that is compatible with recent experimental fabrication techniques, which uses  PCs comprised of periodic arrays of  semiconductor NWs.  The basic ability of NW arrays to form lossless PC waveguides is understood \cite{Johnson2001}, and previous rudimentary fabricated structures have experimentally demonstrated their ability to guide light \cite{Assefa2004,Tokushima2004}.  However, to our knowledge there has not been a proposal to embed QDs in these PC structures. As we will show, the NW PC  structures offer high Purcell factors, $\text{\em F}_d$, high  $\beta$ factors, and have the ability to help overcome the fabrication issues inherent to the traditional PC slab platform.   
We investigate the fundamental physics of these PC NW devices, focusing on the design and optimization of realistic waveguides for QD-enabled light sources, such as is shown in Fig.~\ref{pics}\subref{realwgpic}.   In particular, we propose a novel design for a ``photon gun'', a directed single photon source.  We exploit rigorous Bloch mode analysis and photonic Green function theory to develop a comprehensive model of the behavior of these devices, and demonstrate some of their unique properties and potential advantages over traditional  PC slab structures. 

Our paper is organized as follows.  In Sec. \ref{sec:theory}  we present the  electromagnetic theoretical techniques used to model the PC NW waveguide structures, combining Bloch mode theory and the photonic Green function. The Green function directly connects to the LDOS, photon transport, and the medium-dependent Lamb shift.  We describe the computational techniques used to obtain the optical modes, the Green function of a finite-size PC array, and the underlying photonic band structure.  In Sec. \ref{sec:design} we present the properties of, and explain the physics behind, PC NW waveguides optimized for single photon source applications, focusing on $\text{\em F}_d$ and the $\beta$ factors near standard telecom wavelengths (1550\,nm).  Both infinite and realistic finite-size  structures are studied, and their differences, as well as their performance relative to the state-of-the-art PC slab waveguides, are  discussed.

\section{Theoretical Formalism}
\label{sec:theory}
\subsection{Waveguide modes, photon Green function, Lamb shift and Purcell factor}
 Light propagation through an arbitrary dielectric medium can be described in terms of the mode solutions to the Helmholtz equation: \begin{equation}
\nabla \times \nabla \times \mathbf{f}_{\lambda}(\bold{r}) - \frac{\omega_{\lambda}^2}{c^2} \epsilon(\mathbf{r})\mathbf{f}_\lambda(\mathbf{r})=0,
\label{eq:helmholtz}
\end{equation}   
where $\epsilon(\bold{r})$ describes the relative permittivity of the structure and $\mathbf{f}_\lambda(\mathbf{r})$ are generalized field modes with harmonic $e^{-i\omega t}$ time dependence.  The electric-field Green function, which describes the field response at  $\mathbf{r}$ to a point source at $\mathbf{r'}$ is defined through
\begin{equation}
\left [\nabla \times \nabla \times - \frac{\omega^2}{c^2} \epsilon(\mathbf{r})\\ \right ]\mathbf{G}(\mathbf{r},\mathbf{r}',\omega) =\frac{\omega^2}{c^2}{\mathbf{1}} \delta(\mathbf{r}-\mathbf{r'}),
\label{eq:greens}
\end{equation}
 where ${\mathbf G}_{i,j}$ is a second rank tensor and $\mathbf{1}$ is the unit dyad; element [$i, j$] corresponds to the response in direction $i$ at $\mathbf{r}$ from the $j$th component of the source at $\mathbf{r'}$.  Once one has determined the medium Green function, the field response to an arbitrary polarization dipole source  $\mathbf{P}(\mathbf{r}, \omega)$ can be found from
\begin{align}
\mathbf{E}(\mathbf{r}, \omega)&=\mathbf{E}^{\rm h}(\mathbf{r}, w)
\nonumber \\
&+\frac{1}{\epsilon_o}\int_{V'} \mathbf{G}(\mathbf{r}, \mathbf{r'}; \omega)\cdot \mathbf{P}(\mathbf{r'}, \omega)d\mathbf{r'},
\label{eq:gdef}
\end{align} 
in which $\mathbf{E}^{\rm h}$ is the homogeneous  field solution in the absence of the polarization source.  The eigenmodes
of Eq.~\eqref{eq:helmholtz}, $\mathbf{f}_\lambda(\mathbf{r})$,  form an orthonormal  and  complete set,
so that 
$\int_V \epsilon(\mathbf{r})\mathbf{f}_\lambda(\mathbf{r})\cdot \mathbf{f}_{\lambda'}^*(\mathbf{r})d\mathbf{r} =\delta_{\lambda, \lambda'}$
and 
 $\sum_ {\lambda}\epsilon(\mathbf{r})\mathbf{f}_{\lambda}(\mathbf{r}) \mathbf{f}_{\lambda}^*(\mathbf{r'}) ={\mathbf{1}} \delta(\mathbf{r}-\mathbf{r'})$\cite{Joannopoulos2011, Sakoda2005}, 
where an outer product $\otimes$ is implied for the vector product and the sum also includes longitudinal modes, with $\nabla \cdot \mathbf{D}\!\neq\!0 , \omega_\lambda\!=\!0$ (${\bf D}$ is the displacement field).  
These relationships allow us to write the Green tensor in terms of an expansion over the eigenmodes \cite{Sakoda2005},
\begin{align}
\mathbf{G}(\mathbf{r}, \mathbf{r'}, \omega)=&\sum_{\lambda} \frac{\omega^2}{\omega_{\lambda}^2 - \omega^2}\mathbf{f}_{\lambda}(\mathbf{r}) \mathbf{f}^{*}_{\lambda}(\mathbf{r'}) \nonumber \\ 
&-\frac{{\mathbf{1}} \delta(\mathbf{r}-\mathbf{r'})}{\epsilon(\mathbf{r})}.
\label{eq:Gexp}
\end{align}
In the above, all $\mathbf{f}_\lambda(\mathbf{r})$ are quasi-transverse with $\mathbf{D}\!=\!0 , \omega_\lambda\!\neq\!0$, as they are solutions to Eq.~\eqref{eq:helmholtz}. 

Photonic crystal slabs have discrete translational symmetry in their in-plane dielectric structure, allowing one to employ Bloch's theorem and express solutions as Bloch waves.  Eigenmodes are found to lie in continuous bands, defined by the band-specific dispersion (an $\omega-\mathbf{k}_{\parallel}$ relationship, where $\mathbf{k}_{\parallel}$ is the in-plane wave vector), analogous to electron bands in semiconductors.  In PC waveguides, the presence of a linear defect introduces a localized waveguide band into the bandgap of the surrounding structure \cite{Joannopoulos2011}.  Waveguide modes below the light line ($\omega\!=\!c|\mathbf{k}|$) will propagate without loss through an ideal structure (in the absence of imperfections) and can be written as $\mathbf{f}_{k_\omega}(\mathbf{r})=\sqrt{\frac{a}{L}}\mathbf{e}_{k_\omega}(\mathbf{r})e^{i k_\omega\mathbf{x} }$, where $\mathbf{e}_{ k_\omega}(\mathbf{r})$ is the Bloch waveform, sharing the same periodicity as the lattice, $a$ is the pitch of the PC, and $L$ is the length of the structure; $\mathbf{e}_{ k_\omega}(\mathbf{r})$ is normalized according to $\int_{V_c} \epsilon(\mathbf{r})|\mathbf{e}_{ k_\omega}(\mathbf{r})|^2=1$, where $V_c$ is the spatial volume of a PC unit-cell. These normalized Bloch modes can be substituted into Eq.~\eqref{eq:Gexp}, and by replacing the sum with an integral and carrying out complex pole integration \cite{Yao2009}, one arrives at an analytic expression for the waveguide Green function,
with
\begin{align}
\mathbf{G}_{\rm w}&(\mathbf{r}, \mathbf{r'}, \omega)=\nonumber \\ \frac{i a \omega}{2 v_g}\Big[ &\Theta(x-x')\mathbf{e}_{k_\omega}(\mathbf{r})\mathbf{e}^*_{ k_\omega}(\mathbf{r}')e^{i k_\omega (x-x') } \nonumber \\
+&\Theta(x'-x)\mathbf{e}^*_{k_\omega}(\mathbf{r})\mathbf{e}_{ k_\omega}(\mathbf{r}')e^{i k_\omega (x'-x) } \Big],
\label{eq:Gwg}
\end{align}
where the terms preceded by the first Heaviside function correspond to forward and backwards propagating modes, respectively, and $v_g=|v_g(\omega)|$ is the group velocity at the frequency on interest. By assuming the waveguide mode dominates the optical response throughout the relevant frequency range such that $\mathbf{G} \approx \mathbf{G}_{\rm w}$ (an approximation validated in Sec. \ref{subsec:designideal}), we obtain an analytic expression for the system Green function of a PC waveguide in terms of the Bloch modes, which are readily solvable using the numerical techniques discussed in Sec. \ref{subsec:comp}.

The Green function is a powerful tool in the analysis of  the behavior of a QD, or any two-level emitter, in an arbitrary material system such as a PC waveguide.  In the weak to intermediate coupling regime,  the Lamb shift and spontaneous emission rate of an emitter are given by
(e.g., see Refs.~\onlinecite{Yao2009,Dung2002, Wubs2004}):
\begin{equation}
\Delta \omega=-\frac{1}{\hbar\epsilon_0}\mathbf{d} \cdot\text{Re}\left\{ \mathbf{G} ( \mathbf{r}_d,\mathbf{r}_d\omega_{d})\right\}\cdot \mathbf{d},
\label{eq:dw}
\end{equation}
and
\begin{equation}
 \Gamma=\frac{2}{\hbar\epsilon_0}\mathbf{d}\cdot\text{Im}\left\{ \mathbf{G} ( \mathbf{r}_d, \mathbf{r}_{d}; \omega_{d})\right\}\cdot \mathbf{d},
\label{eq:A}
\end{equation}
where $\mathbf{d}$ is the dipole moment of the emitter at position $\mathbf{r}_d$.  From  Eq.~\eqref{eq:A} the spontaneous emission rate, or Einstein A coefficient, of a single QD is directly proportional to $\text{Im}\left\{ \mathbf{G} ( \mathbf{r}_d, \mathbf{r}_d; \omega_{d})\right\} $, projected along the direction of the dipole\cite{Wubs2004}.  Thus, we define the enhanced spontaneous emission (or generalized Purcell) factor as
\begin{equation}F_d(\mathbf{r}_d, \omega)=\frac{\text{Im}\{\hat {{\bf n}}_d \cdot \mathbf{G}(\mathbf{r}_d, \mathbf{r}_d; \omega)\cdot {\hat{\bf n}}_d\}}{\text{Im}
\{{\hat{\bf n}}_d \cdot\mathbf{G}^{\rm h}(\mathbf{r}_d, \mathbf{r}_d; \omega)\cdot {\hat{\bf n}}_d\}}.
\label{eq:pf}
\end{equation}
In the above, ${\hat{\bf n}}_d $ is a unit vector along the photon emitter's dipole moment. Using the  known Green function of a homogeneous medium with the dielectric constant $\epsilon_h$: $\text{Im}\{{\hat{\bf n}}_d \cdot\mathbf{G}^{\rm h}(\mathbf{r}_d, \mathbf{r}_d, \omega)\cdot {\hat{\bf n}}_d\}= \omega^3 n^{\rm h}_d/(6\pi c^3)$\cite{Novotny2006}, together with Eq.~\eqref{eq:Gwg}, we  obtain  an analytic expression for the Purcell factor of a single QD in a PC waveguide:
\begin{equation}
F_d(\mathbf{r}_d, w)=\frac{3\pi a c^3}{n^{\rm h}_d \omega^2 v_g}
\left |\mathbf{e}_{k_\omega}(\mathbf{r}_d)\cdot \mathbf{\hat{n}}_d
\right |^2,
\label{eq:pfa}
\end{equation}
where $n^{\rm h}_d=\sqrt{\epsilon^{\rm h}_d}$, the relative permittivity of the background medium at the QD's location.  At a field antinode location ${\bf r}_0$, with perfect polarization coupling,    Eq.~\eqref{eq:pfa} can be written in terms of the familiar Purcell factor expression typically applied to cavities\cite{Rao2007theory}
\begin{equation}
\text{P}_{\rm F} = \frac{3}{4\pi^2}\left(\frac{\lambda}{n^{\rm h}_d}\right)^3 \left (\frac{Q}{V_{\rm eff}} \right),
\label{eq:PF}
\end{equation}
where  the effective mode volume, per unit cell, $V_{\rm eff}=1/\epsilon^{\rm h}_d|\mathbf{e}_{k_\omega}(\mathbf{r}_0)|^2$.  The quality factor $Q=\omega/\Gamma_c$, where $\Gamma_c=2v_g/a$ is the effective ``open-cavity decay rate'' of the PC waveguide and represents the decay rate into the   waveguide mode. 

\subsection{Computational techniques for obtaining the optical modes and Green functions}
\label{subsec:comp}
To model the classical light-matter interaction of PC NW arrays, we use several different computational tools and methods.  First, the open source plane-wave expansion software MIT Photonic Bands (MPB) \cite{Johnson2001} was used to calculate the Bloch modes, band structure, group velocity, and theoretical PF of ideal periodic structures.  We then focus our study on finite-size PCs, which cannot be treated with simple Bloch mode analysis, and instead make use of  the more direct finite-difference time-domain (FDTD) approach\cite{Sullivan2013}.  To model finite-sized structures we use the FDTD software developed by Lumerical \cite{LumericalSolutions}; importantly, this finite-size technique allows us to obtain the radiative losses and coupling efficiencies of a single photon emitted into a finite-size waveguide array.  A point polarization dipole is placed at the desired QD position (${\bf r}_d$) and with a polarization that is aligned with the QD dipole and the mode of interest;  by recording the time-dependent electric field, we  calculate a numerically exact expression for the Green function, derived directly from its definition in Eq.~\eqref{eq:gdef}. Defining $\mathcal{F}()$ as the Fourier transform function, then
\begin{equation}
\mathbf{G}(\mathbf{r}, \mathbf{r}_d; \omega)_{i,j}=\frac{\mathcal{F}(\mathbf{E}(\mathbf{r}, t)\cdot \hat{\mathbf{i}} )}{\mathcal{F}(\mathbf{P}(\mathbf{r}_d, t)/\epsilon_0 \cdot \hat{\mathbf{j}})}.
\label{eq:Gfdtd}
\end{equation}The PF can easily be found from Eq.~\eqref{eq:pf}, as can the Lamb shift \footnote{From Eq.~(\ref{eq:Gexp}), the real part of {$\mathbf{G}$} diverges at {$ \mathbf{r}\!=\!\mathbf{r}'$}.  This term only contains a finite (photonic) component from the medium-dependent $\mathbf{G}$, with the divergence originating from the homogeneous Green function ({$\mathbf{G}^{\rm h}$}).  The divergent contribution to the lamb shift is already present in the resonant QD frequency\cite{Sakoda2005}, and a re-normalization technique is employed, subtracting off the simulation-mesh-size-dependent homogeneous Green function contribution to only consider the finite photonic contribution to the Lamb shift.} using Eq. \eqref{eq:dw}. 

A numerical check was preformed to verify the accuracy of this FDTD Green function approach, and it was found to recover the analytic answer for a homogeneous structure with errors of less than 1\% over the entire frequency spectrum initially excited by the dipole (typically $\sim$50\,THz); we have also carried out similar checks for inhomogeneous structures, such as spheres and half space geometries elsewhere. Simulations in our FDTD approach are bounded by perfectly matched layers (PMLs), which allow light to propagate out of the computational structure.  This allows for the treatment of radiative losses, enabling the calculation of band structures above the light line, finite-size mode profiles, and the $\beta$ factor. The FDTD Green function approach thus allows us to determine both the real and imaginary parts of the complex eigenmodes.  Using spectral filtering and apodization to remove source effects, we calculate mode profiles $\mathbf{E}_\lambda$ from the full FDTD eigenmodes.  Here  we define the $\beta$ factor as the fraction of the power emitted by the simulation dipole which exits the waveguide via the waveguide channel in the desired direction: $\beta=P_{\rm wg}/P_{\rm source}$.  Power flow is calculated through a surface integral of the Poynting vector $\mathbf{S}$ such that $P_{\rm wg}={\int_S}_{\rm wg} \mathbf{S} \cdot d\mathbf{a}$ and $P_{\rm source}={\oint_S}_{\rm tot} \mathbf{S} \cdot d\mathbf{a}$ where $S_{\rm wg}$ is two planes, at either end of the structure, normal to the propagation direction, and $S_{\rm tot}$ is a box bounding the entire structure.  

\section{Waveguide Design}
\label{sec:design}
\subsection{Band structure and spontaneous emission enhancements in NW waveguides}
\label{subsec:designideal}
 \begin{figure}[h!]
\vspace{-5pt}
\subfloat[]{\label{topscheme}\includegraphics[width=0.49\textwidth]{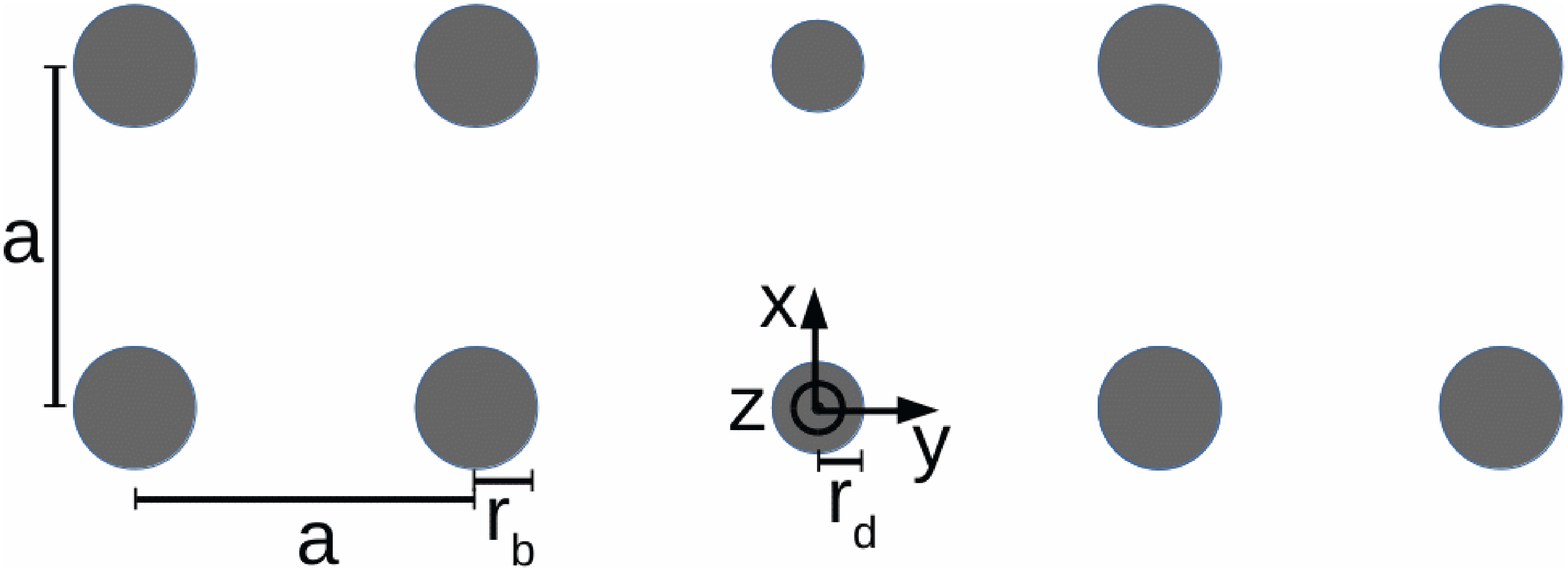}}

\vspace{-5pt}
\subfloat[]{\label{bs1}\includegraphics[width=0.49\textwidth]{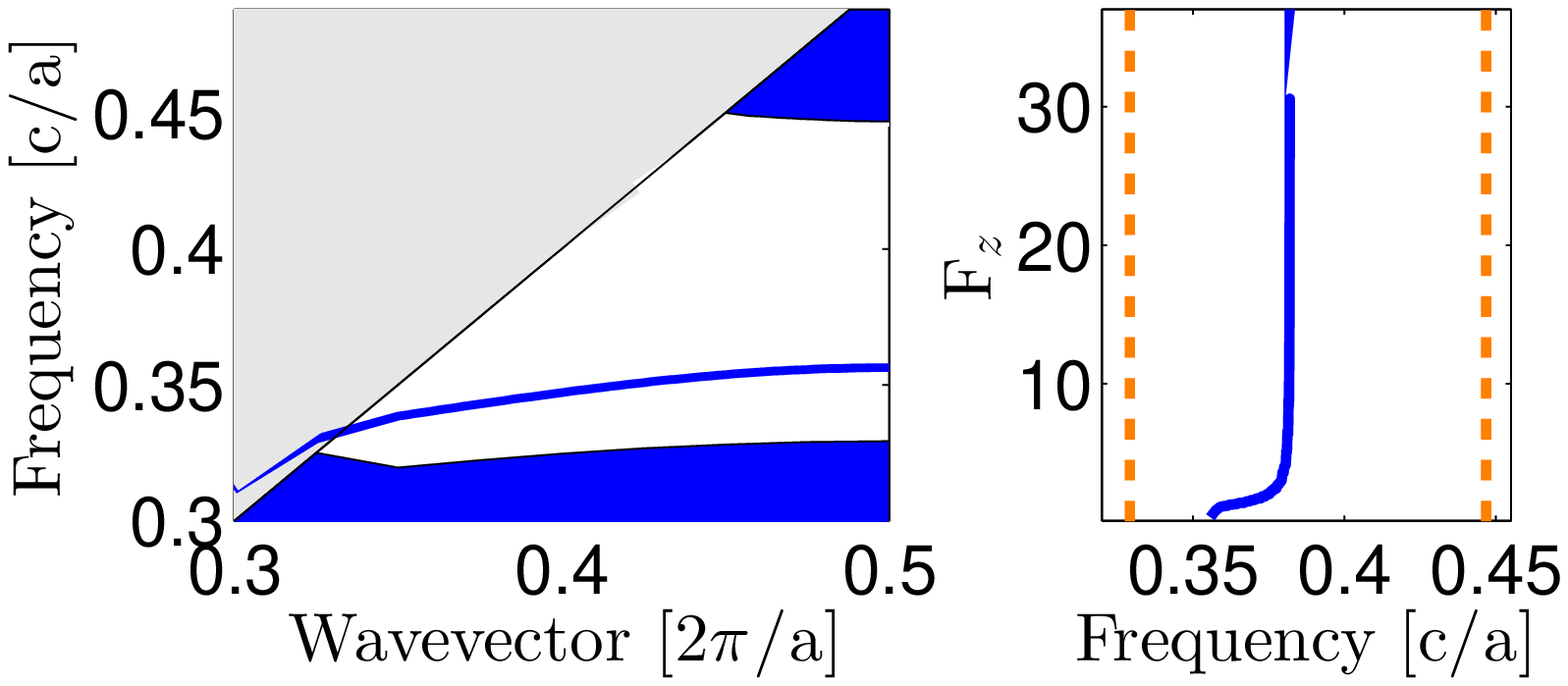}}

\vspace{-5pt}
\subfloat[$f\!\!=\!\! 0.3569\frac{c}{a}$ \vspace{-2pt}]{\label{blochprop}\includegraphics[width=0.49\textwidth]{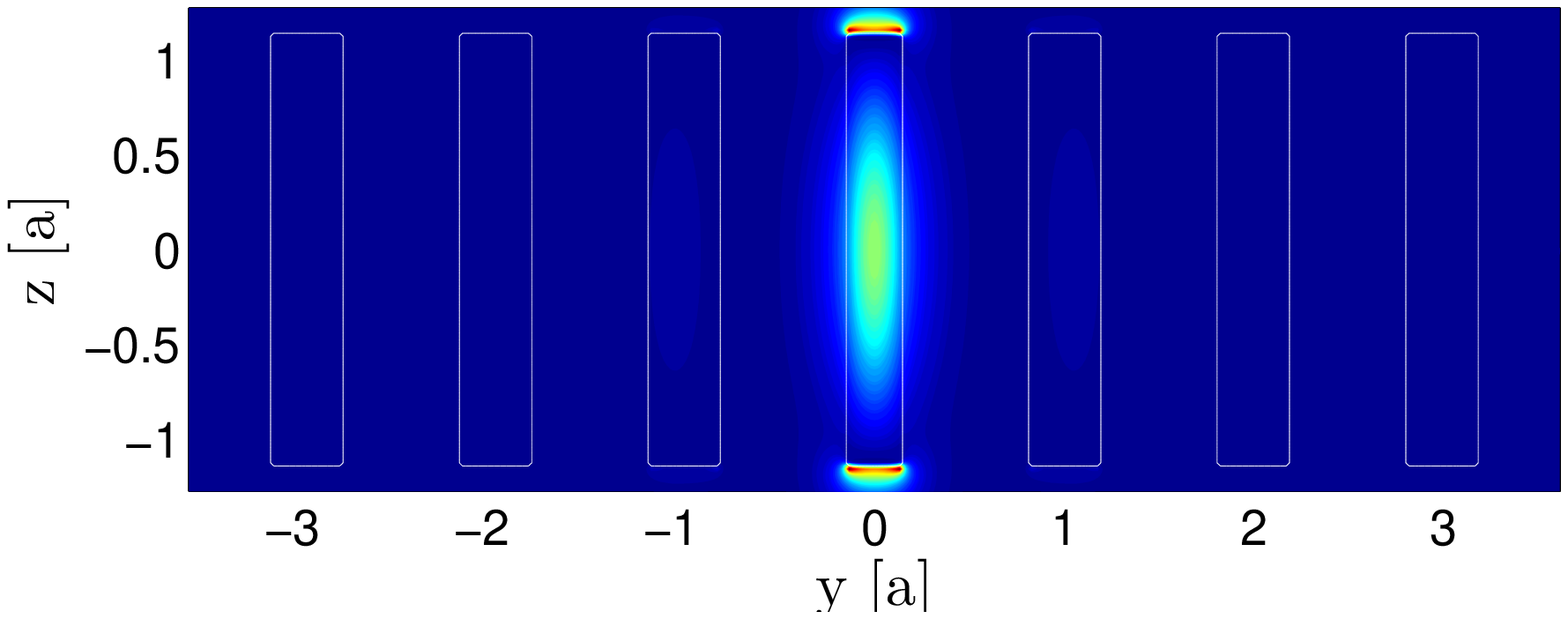}}
\vspace{-5pt}
\caption{{(Color online) (a) Top view schematic of two unit cells of a general NW waveguide, showing  structural parameters and coordinate convention used throughout this paper.  (b) Band structure on left and spontaneous emission enhancement factor on right, as calculated with MPB, of a NW waveguide with $r_d=0.14a$.  The bulk PC bands are shaded in blue, and the region above the light line in gray, with the guided band  lying in the band gap.  $F_z$ is calculated for an emitter in the center of the waveguide NW, and can be seen to diverge as one approaches the mode edge.   The band gap is bounded with dotted orange lines. (c) Profile  $|\mathbf{E}_\lambda|^2$, in arbitrary units, of the indicated waveguide mode perpendicular to the waveguide direction.  The monopole-like profile and strong confinement to the waveguide channel are clearly visible. }
\label{intro}}
\vspace{-10pt}
\end{figure}
In a PC NW array, one can produce a waveguide by reducing the radius of a single row of NWs, as shown schematically in Fig.~\ref{intro}\subref{topscheme}.   Nanowire PC arrays contain a photonic bandgap between the first and second odd (or TM-like, as they share the properties of TM modes in pure two-dimensional structures) bands\cite{Joannopoulos2011}.  The lowest  order band has a monopole-like mode profile, with most of the field energy localized to the NW.  By reducing the radius of a row of NWs, one decreases the effective index seen by a mode propagating along this channel, producing a waveguide band by blue-shifting the lowest order band into the surrounding photonic bandgap\cite{Johnson2000}.  Modes in this waveguide band will decay evanescently away from this waveguide channel, as they lie in the bandgap of the surrounding structure.

 The band structure of a NW waveguide can be seen in Fig.~\ref{intro}\subref{bs1}, and  a Bloch mode profile perpendicular to the waveguide direction is shown in Fig.~\ref{intro}\subref{blochprop}.   We note that results are given in scale invariant units, as the operating frequency of these structures can be adjusted  by simply tuning the pitch.  Because the lowest-order band is being pulled up into the band gap, the LDOS, and resulting spontaneous emission factor of the guided modes increase with frequency up to the mode edge, which imposes a high frequency cut-off, as seen in Fig.~\ref{intro}\subref{bs1}.  This result is in contrast with the traditional slab PC waveguide, which contains an even guided band with a low frequency mode edge \cite{Johnson2000}.  As the waveguide mode is odd, the relevant mode profile $\mathbf{E}_\lambda \approx E_z(\omega_\lambda)$ and we plot $|{\bf E}_\lambda|^2$ as it is directly proportional to the Green function at equal space points, as per Eq. \eqref{eq:Gexp}.

 \begin{figure}[h!]
\vspace{-5pt}
\subfloat[\vspace{-2pt}]{\label{struccomp}\includegraphics[width=0.49\textwidth]{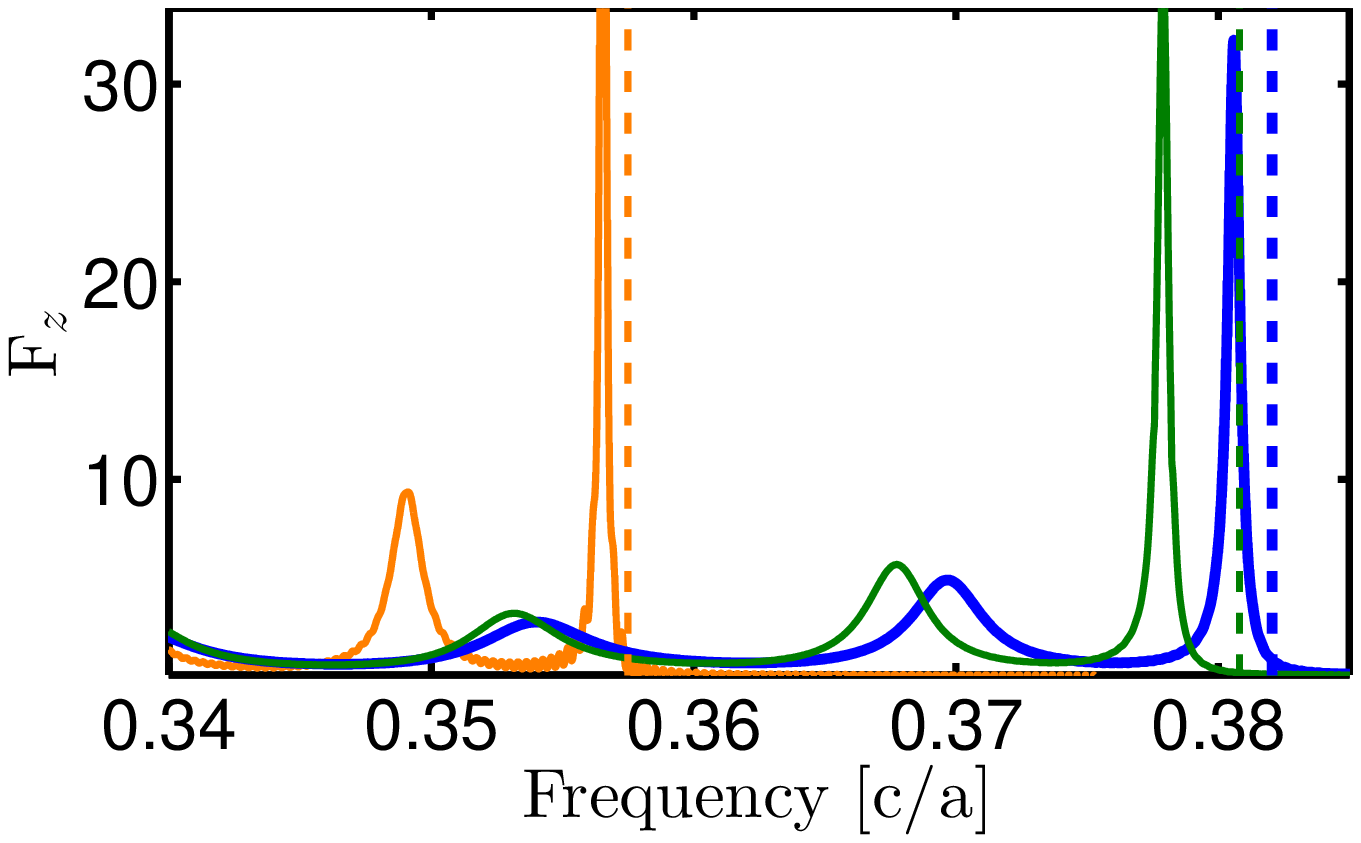}}

\vspace{-5pt}
\subfloat[\vspace{-2pt}]{\label{bs1}\includegraphics[width=0.49\textwidth]{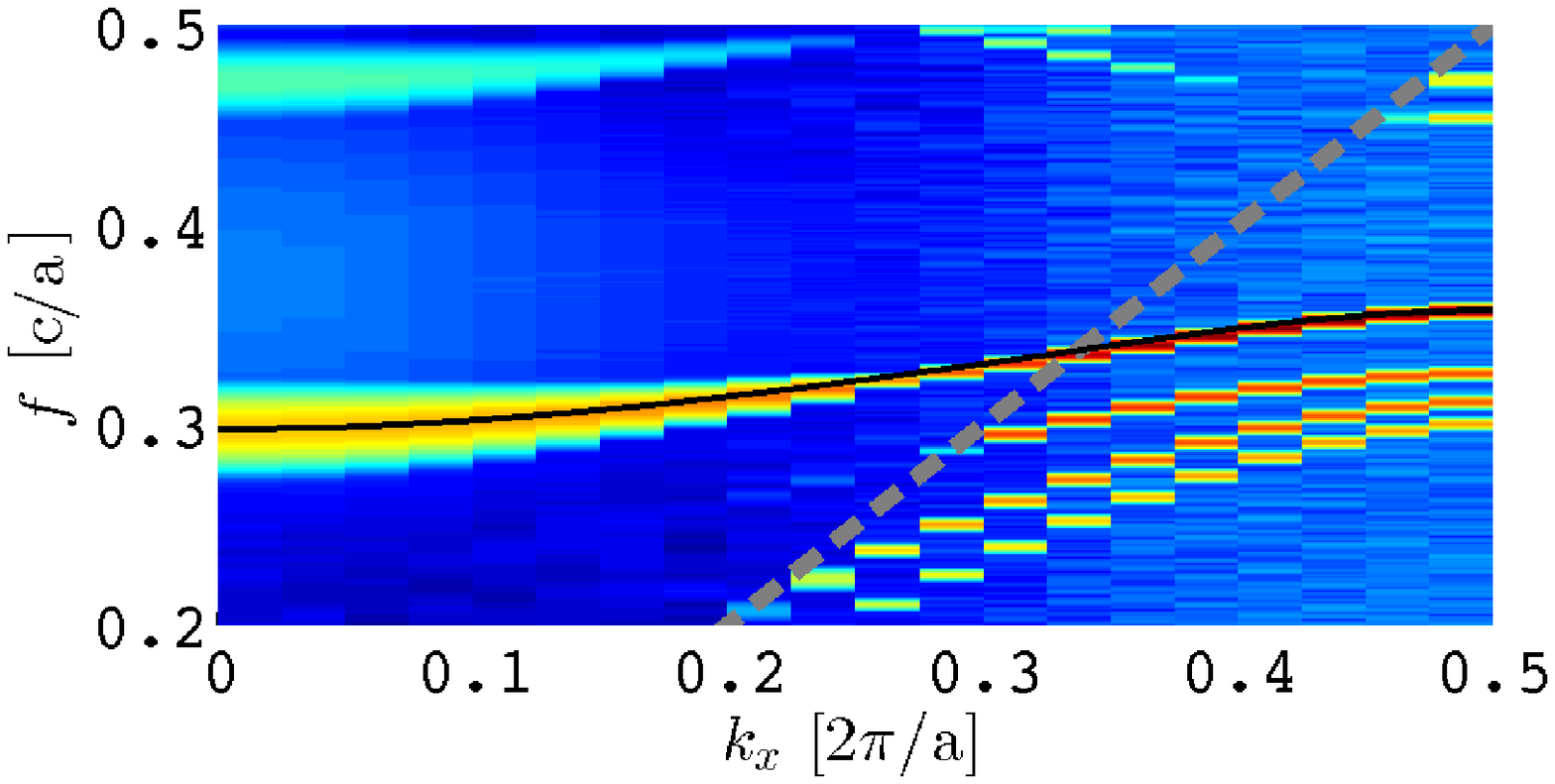}}

\vspace{-5pt}
\caption{(Color online) (a) Spontaneous emission enhancement factors for QDs embedded in the center of the finite-size waveguide NW  PC with length 15$a$.    Thick (dark) blue indicates heterogeneous design, and (light) orange and (dark) green correspond to thick and thin homogeneous NWs.  Dotted lines indicate the mode edges computed for corresponding infinite structures in MPB.  (b) Band structure from FDTD  of homogeneous NW waveguide with $r_d=0.140a$.  The light line is indicated in grey, waveguide band highlighted in black, and modal strength is shown on a logarithmic scale. Above the light line, the waveguide band  broadens, as it couples to radiation modes and becomes leaky. }
\label{design}
\vspace{-15pt}
\end{figure}Using MPB, both heterogeneous and homogeneous PC structures comprised of NWs suspended in air were considered.  Homogeneous NWs had a dielectric constant $\epsilon=13$, while heterogeneous NWs had alternating layers of $\epsilon=13$ and $\epsilon=12$ arranged in a Distributed Bragg Reflector (DBR) pattern.  Structural parameters were chosen to optimize the PC array band gap for both design types, with homogeneous NWs having a radius of $r_b=0.180a$ and height  $h=2.27a$, and heterogeneous having  $r_b=0.189a$ and $h=2.10a$.  Both structures have a square lattice structure as this was found to yield a larger slow-light region in the guided band of these waveguides, with all NWs 1$a$ apart.  The waveguide NW radius ($r_d$) was tuned to localize the waveguide band in the center of the surrounding bandgap, resulting in $r_d=0.120a$ and  $r_d=0.130a$ for homogeneous and heterogeneous structures, respectively.  We also considered an alternative design for the homogeneous NW structure where the flatness of the waveguide band, as opposed to the location of the mode edge, was maximized, resulting in $r_d=0.140a$;  Fig.~\ref{intro}\subref{topscheme} is dimensioned according to this design.  All three designs used a pitch of $a=0.5655\,\mu m$, designed to have a fundamental waveguide mode edge near the standard telecom wavelength of $1550\,$nm.

   The enhanced emission factor,  $F_z$, of a (vertically polarized) QD in the center of a NW for all three structures was calculated for an infinite PC structure using MPB, and in the central NW of a 15$a$ long waveguide using FDTD and Eqs.~\eqref{eq:Gfdtd} and \eqref{eq:pf}; see Fig.~\ref{design}\subref{struccomp}.  In both MPB and FDTD, waveguide widths of 7$a$ were used, corresponding to three rows of background PC NWs on either side of the waveguide array.  This was found to be sufficient to almost entirely eliminate in-plane losses in those directions, demonstrating the utility of  PC physics.  The finite-size structures  are truncated abruptly and surrounded by a substantial volume of free space before the termination of the simulation volume with PML to prevent clipping.  A number of important finite-size effects can clearly be seen.  Firstly, the DOS no longer divergences at the waveguide mode edge, instead forming a red-shifted strong resonance referred to in this paper as the band edge quasi-mode, ($\lambda_{0}$).  In addition, weaker Fabry-P\'{e}rot (FP) quasi-mode resonances ($\lambda_{{\rm FP}}$) can be seen throughout the waveguide band, arising from reflections off the waveguides' terminus.  Similar $\beta$ factors were determined for all three designs, with values in the 88-90\% range throughout the waveguide band, increasing to $\sim$95\% at FP resonances and $\sim$98\% at $\lambda_{0}$.  We highlight that ($i$) these $\beta$ factors exceed those  in cutting-edge PC slab waveguides \cite{Lund-Hansen2008}, and that these ($ii$) finite-size effects are both predicted \cite{Rao2007} and seen experimentally \cite{BaHoang2012} in slab PC waveguides as well.  Superior $\beta$ factors are obtained due to the waveguide modes being vertically polarized in NW PC structures, minimizing out-of-plane losses, while in-plane losses are almost entirely eliminated by the surrounding PC layers. .

From examining  Fig.~\ref{design}\subref{struccomp}, it is evident that the heterostructured NWs show little improvement in single photon properties over their homogeneous counterparts, as the index contrast is too weak and the NWs too short for the DBR layers to have a noticeable effect.  We note that as the length of the NW increases, the bandgap begins to shrink as it becomes easier to add additional vertical nodes to form higher order modes, limiting the range of NW lengths available\cite{Johnson1999}.  The larger waveguide radius design generates a higher peak $F_d$ when finite-sized structures are considered, without any compromise in $\beta$ factor, and thus was chosen and is used in all subsequent structures.  The band structure of this design computed in FDTD along the waveguide ($x$) direction can be seen in Fig.~\ref{design}\subref{bs1}, with a strong and flat waveguide band is clearly visible in the surrounding bandgap. This dispersion flatness is likely a result of the frequency at the edge of the first Brilloiun zone ($\mathbf{k}=0.5\,\frac{\pi}{a}$) being tuned to be close to the waveguide band frequency at $\mathbf{k}=0$, which is independent of NW radius.  The FDTD band structure shows agreement with that of Fig.~\ref{intro}\subref{bs1} (MPB) below the light line as expected, but is also valid above the light line from the inclusion of radiative loss.  This treatment of radiative losses in FDTD allowed the low $\mathbf{k}$ band structure to be understood and NW waveguide radius optimized to produce this design.  As is evident from Eq.~\eqref{eq:pfa}, the low $v_g$ throughout the guided band is the source of the increased $F_d$.

\subsection{Finite-size NW waveguides}
\label{subsec:idealfinite}
\begin{figure}
\vspace{-5pt}
\subfloat[\vspace{-2pt}]{\label{lengthpfb}\includegraphics[width=0.49\textwidth]{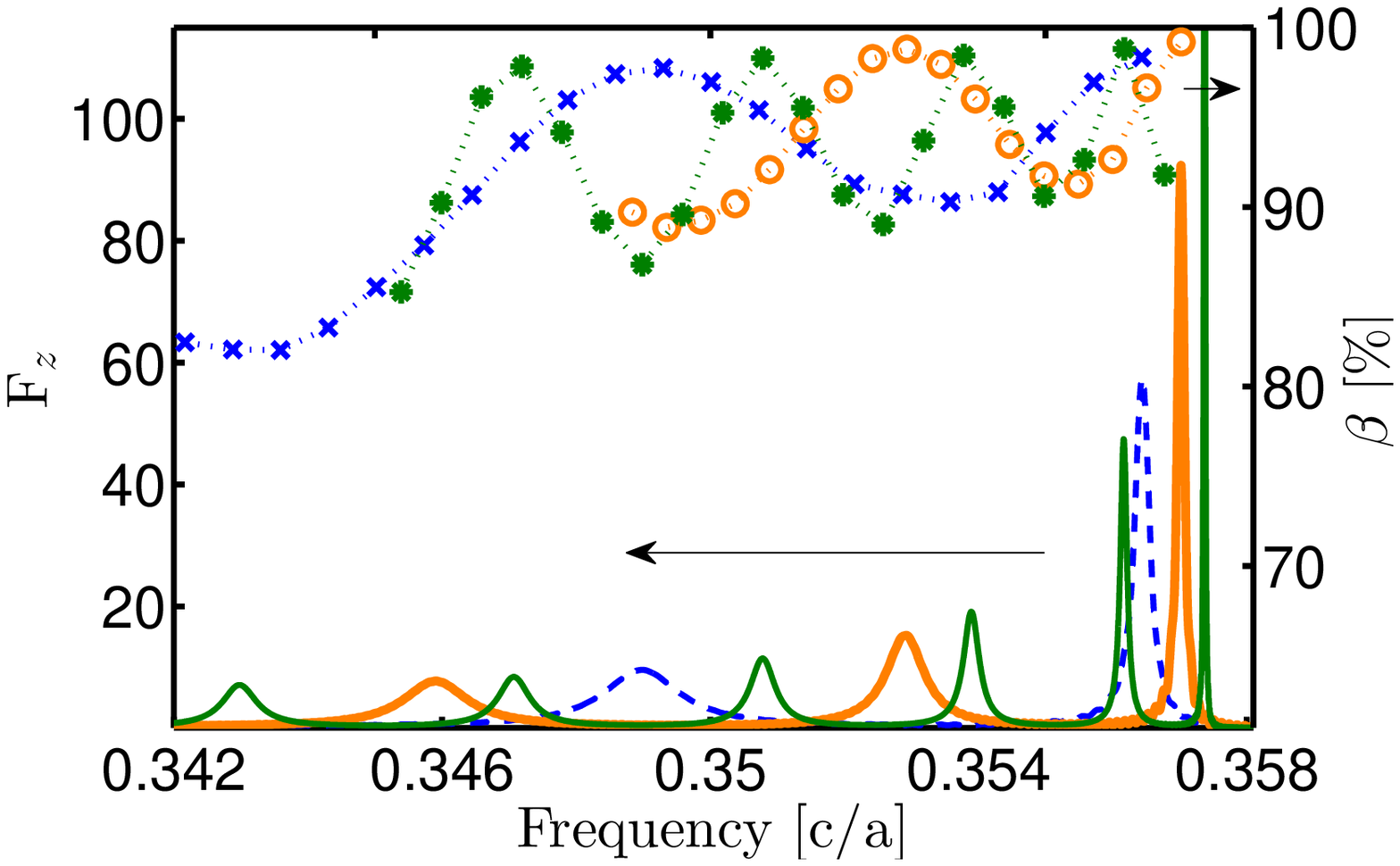}}

\vspace{-5pt}
\subfloat[$f\!\!=\!\!0.3529\frac{c}{a}$\vspace{-2pt}]{\label{fprip}\includegraphics[width=0.49\textwidth]{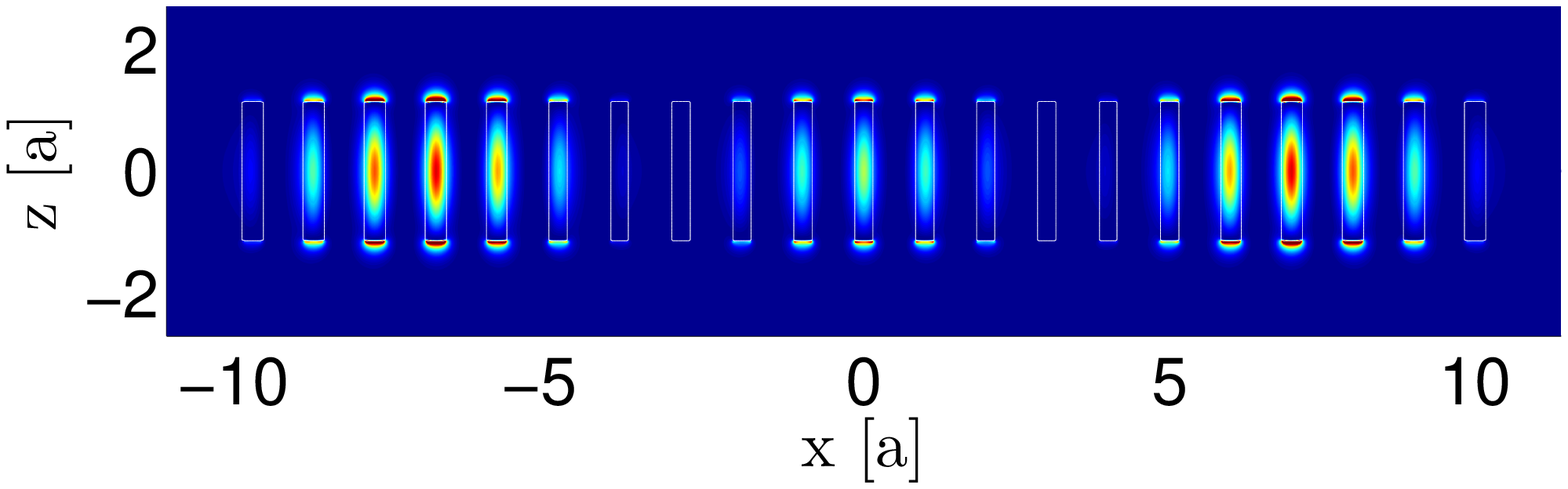}}

\vspace{-5pt}
\subfloat[$f\!\!=\!\!0.3570\frac{c}{a}$\vspace{-2pt}]{\label{bedge}\includegraphics[width=0.49\textwidth]{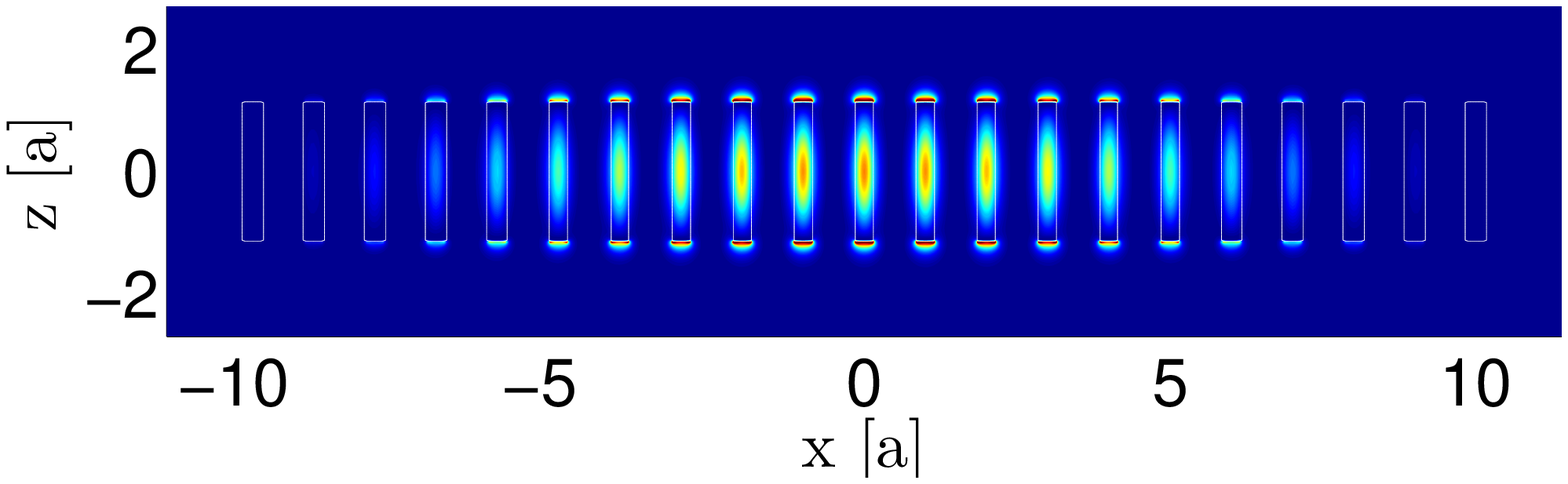}}
\vspace{-5pt}
\caption{(Color online) (a) Purcell factors and $\beta$ factors of a finite-size NW PC waveguide, for various waveguide lengths as a function of increasing length;  Purcell ($\beta$ factor) for 15$a$ waveguide in blue dashed line (`$\times$'), 21$a$ in light orange (`$\circ$'), and 41$a$ in dark green (`$\ast$').  $\beta$ factors are calculated at discrete frequency points, and the dotted lines are provided only to guide the eye.  The 41$a$ $\lambda_{0}$ resonance peak $F_z$ increases to 384, although the axis terminates at 120.  (b) and (c):  $|{\bf E}_\lambda|^2$, in arbitrary units for $\lambda=\lambda_{{\rm FP}}$ in (b) and $\lambda=\lambda_0$ in (c) on the $y=0$ plane of the 21$a$ waveguide.}
\label{length}
\vspace{-15pt}
\end{figure}Figure~\ref{length}\subref{lengthpfb} explores finite-size effects in more detail, comparing the $F_d$ and $\beta$ factors of an emitter in the center of the central NW of 15$a$, 21$a$, and 41$a$ length waveguides of the chosen design.  The pitch has been reduced to $a=0.5526\,\mu m$ to shift the mode edge closer to the $1550\,$nm range.  It can be seen that with increasing waveguide length, the mode edge quasi-mode narrows, blue shifts, and its peak value increases substantially as it comes closer to the result found for the infinite structure.    In addition, the number of FP resonances increases.  Similar effects are seen and understood for PC slab waveguides \cite{Rao2007, BaHoang2012}.  Mode edge $F_z$ of 57.5, 92.4, and 384 with corresponding  $Qs$ of 1348, 2960, and 23550 at $f_0= 0.3564$, 0.3570, and 0.3574$\,\frac{c}{a}$ are calculated for 15$a$, 21$a$, and 41$a$ waveguides, respectively.  The $f_{0}$ resonance for the 41$a$ waveguide has thus effectively converged to the mode edge of the infinite structure, calculated at 0.3575$\,\frac{c}{a}$.  Very large $\beta$ factors are  clearly seen throughout the guided band, with values of at least 90\%, increasing to the 95-97\% range at FP resonances for all three structures, and 98.3\%, 99.2\%, and 98.8\% at $f_{0}$ for the 15$a$, 21$a$, and 41$a$ waveguides, respectively.  It is also evident for our structures that the $\beta$ factor directly follows the $F_z$, which is advantageous for single photon applications as it allows one to exploit propagating modes with both high emission rate enhancement and high collection efficiency.  Figures~\ref{length}\subref{fprip} and \ref{length}\subref{bedge} show mode profiles of the $21a$ waveguide in a slice through the center of the waveguide array.   The waveguides support a Bloch-like mode which is modulated by the finite-size of the structure, with the field confined tightly to the waveguide NWs.  As $\mathbf{G}$ is directly proportional to the mode profile (Eq.~\eqref{eq:Gexp}), and Figs.~\ref{intro}\subref{blochprop}, \ref{length}\subref{fprip}, and \ref{length}\subref{bedge} indicate that the mode profile in the vicinity of the PC waveguide is entirely dominated by the Bloch mode, the approximation used in deriving  Eq.~\eqref{eq:Gwg} is justified. Figure~\ref{length}\subref{fprip} shows the mode profile corresponding to the strongest FP quasi-mode, and  Fig.~\ref{length}\subref{bedge} shows the mode profile at the band edge resonance.

\subsection{Effects of a substrate and different QD locations}
\label{subsec:real}

Since it is somewhat unrealistic to assume the NWs will that are suspended in air,  we also investigate a number of different substrate designs.  In addition, when the MBE technique is used to embed a QD in the center of a waveguide NW, it will produce an identical QD in every waveguide NW \cite{Makhonin2013}.  While this type of system has the potential to act as a many-body simulator \cite{Hartmann2008}, these additional QDs would serve as a source of loss in a single-photon-source waveguide and lead to poor output coupling. Work with NV centers in diamond has demonstrated deterministic control over emitter position in diamond NWs \cite{Babinec2010}, and the structures considered in this paper can be readily adapted to a diamond NW and NV center base.  Alternatively, one could embed a single QD on top of the central NW, e.g.,  through the fabrication process described in Ref.~\onlinecite{Pattantyus-Abraham2009}.
   The Bloch mode field anti-node is in fact at the edges of the NWs, resulting in an increase in $F_z$ for an emitter on top of a NW relative to the center, making this design advantageous from a performance standpoint as well.  QDs resting on the surface of a slab PC structure has been investigated in Ref.~\onlinecite{Foell2012}, and we use a similar approach here. Note that any index contrast between the QD and the surrounding media will result in a  geometry dependent depolarization, reducing the field seen by the QD.  As this ``Lorentz factor'' \cite{Novotny2006} would have the same strength in an identical homogeneous medium, the depolarization is best thought of as included in the QD dipole moment and has no impact on $F_z$\cite{Foell2012}.

\begin{figure}

\vspace{-5pt}
\subfloat[\vspace{-2pt}]{\label{realcomp}\includegraphics[width=0.49\textwidth]{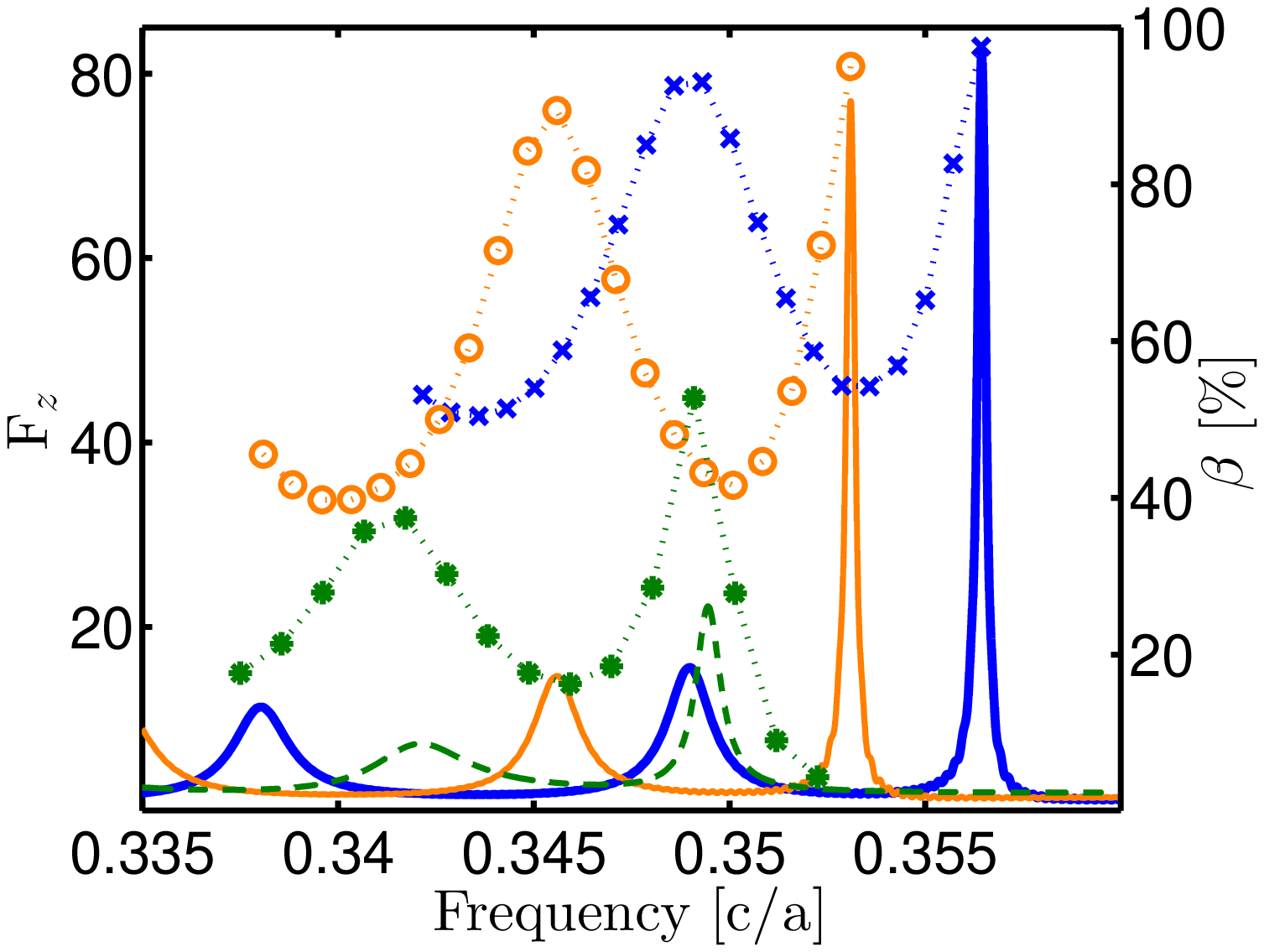}}

\vspace{-5pt}
\subfloat[$f\!\!=\!\! 0.3531\frac{c}{a}$\vspace{-2pt}]{\label{realprof}\includegraphics[width=0.49\textwidth]{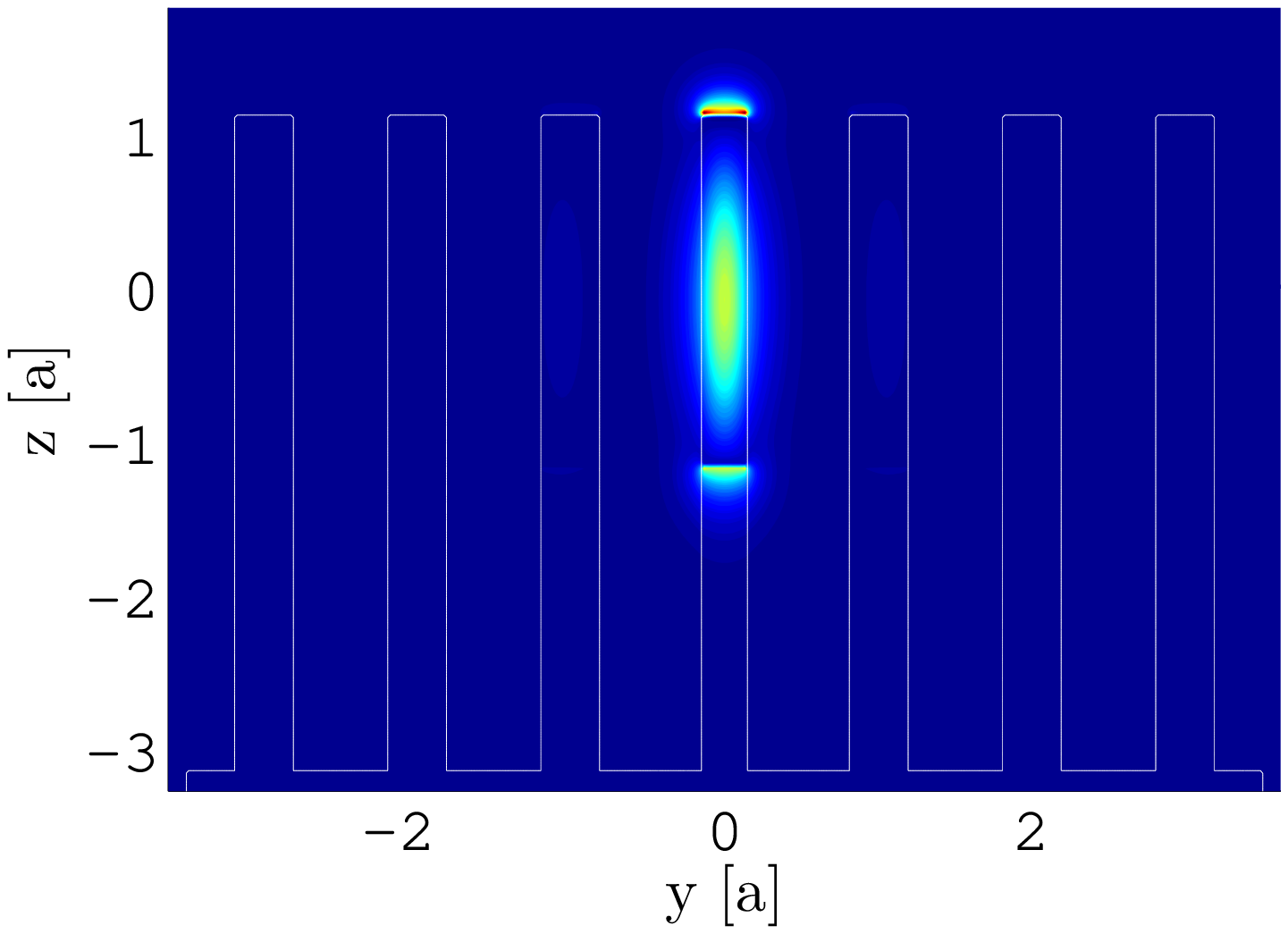}}
\vspace{-10pt}
\caption{(Color online) (a) Purcell and $\beta$ factors of 15$\,a$ waveguides with top mounted QDs.  Substrate-free structure $F_z$ ($\beta$ factor) in thick blue (`$\times$'), structure with a simple substrate in dashed dark green (`$\ast$') and elevated NW design (described in text) in light orange (`$\circ$'). (b) $|{\bf E}_{\lambda_0}|^2$ in $x=0$ plane of elevated NW waveguide, in arbitrary units. }
\label{real}
\vspace{-15pt}
\end{figure}
Three substrate designs were considered, and the Purcell and $\beta$ factor spectrum of 15$a$-length NW PC waveguides with a top-mounted QD utilizing two of these designs, alongside a substrate-free waveguide, are shown in Fig.~\ref{real}\subref{realcomp}.  We first considered a simple substrate directly below the waveguide NWs, corresponding to the dashed line and `$\ast$' symbols.  One can see that this substrate is a large source of loss, yielding low $\beta$ factors, and by breaking the symmetry of the structure drastically reduces the strength and confinement of the waveguide mode, causing a substantial drop in $F_d$.  In order to preserve symmetry, we then considered encasing the structure in a lower index material, such as the structure studied in Ref.~\onlinecite{Tokushima2004} (Si rods, $n=3.48$, in SiO$_2$ and polymer, $n=1.45$).  Structures were modeled using MPB with a background index ranging from $n_{b}=1.1-2$.  In all cases, the reduced the index contrast between the NWs and the surrounding medium decreased the size of the surrounding bandgap, leading to large in-plane losses and a weak guided band with a low $F_z$ for $n_{b}>1.2$.

  In our final design, the PC NW array was extended using a low-index material (AlO, $\epsilon=3.1$), which terminated in a substrate of the same material.  A schematic depiction of this structure is shown in Fig.~\ref{pics}\subref{realwgpic}, and we note that a similar waveguide design was originally proposed in Ref.~\onlinecite{Johnson2000} and implemented in Ref.~\onlinecite{Assefa2004}, who were able to produce the AlO layer by first growing an AlAs layer using MBE and then using a wet thermal oxidation process.  Simulations in FDTD and MPB indicated that guiding was achieved entirely in the high index upper portion of the NWs, with the lower AlO section separating the PC structure from the substrate and dramatically reducing its symmetry-breaking effects.  An AlO NW height of $2a$ was found to be sufficient to eliminate most of the detrimental effects of the substrate, and the properties of this structure are shown in Fig.~\ref{real}\subref{realcomp}, and a mode profile perpendicular to the waveguide direction in Fig.~\ref{real}\subref{realprof}.  The $F_z$, Q, and $f_0$ of the quasi-mode are 77.1, 1282, and $0.3531\frac{c}{a}$,  comparable with values of 82.2, 1332, and $0.3564\frac{c}{a}$ for the substrate-free structure (both with a top-mounted QD), with the red-shift originating from the increase in effective index due to the AlO layer.  The $\beta$ factor for both structures away from any resonances is substantially lower than seen earlier for centrally embedded QDs, as it is far easier for photons not coupling into a waveguide mode to escape vertically.  However, we note that $\beta$ factors as high as 89.4\% and 95.0\% at the largest FP resonance and mode edge respectively are calculated for the realistic structure (93.0\% and 97.6\% for the substrate-free version), with most of the loss occurring vertically.  The waveguide mode profile of Fig.~\ref{real}\subref{realprof} confirms that the substrate has little qualitative effect, as it is largely identical to that of  Fig.~\ref{intro}\subref{blochprop}, with the light residing in the high index upper portion of the waveguide NW.  The mode profile also demonstrates the large field enhancement directly above the waveguide NW.  Thus, we were able to design producible structures without significant loss in key properties, particularly $F_d$ and $\beta$ factors. For the remainder of this paper, we will study PC structures following this more realistic design, with elevated NWs, a substrate, and top-mounted QDs, unless explicitly stated otherwise.

\subsection{Photonic Lamb Shifts}
\label{subsec:lambshift}
\begin{figure}
\includegraphics[width=\columnwidth]{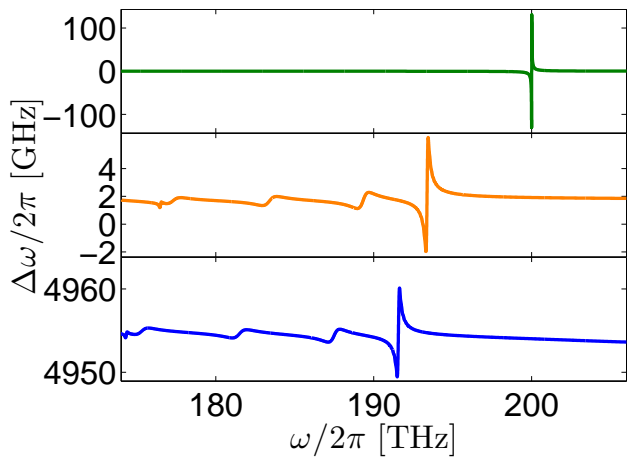}
  \caption{Lamb shift from a  30-debye QD in various PC structures. Top:  QD at antinode of  simple high-Q Lorentzian cavity.  Middle:  QD in center of the 15$a$ substrate-free waveguide.  Bottom:  QD at top of the 15$a$ waveguide with elevated substrate design.}
\vspace{-0pt}
\label{lamb}
\vspace{-10pt}
\end{figure}

Although a good part of this paper has focused on exploiting ${\rm Im}[{\bf  G}({\bf r}_d,{\bf r}_d;\omega_d)]$, 
${\rm Re}[{\bf G}({\bf r}_d,{\bf r}_d;\omega_d)]$ is responsible for the Lamb shift, which is an important and measurable quantum effect that causes a medium-dependent frequency shift of the emitter.  In a simple Lorentzian cavity, the Green tensor is assumed to be single mode, resulting in a simple analytic expression for ${\rm Re}[{\bf G}] \propto 
{\rm Re}[1/(V_{\rm eff}\epsilon(\omega_c^2-\omega^2-i\omega\Gamma_c))]$, and the Lamb shift via Eq.~\eqref{eq:dw}.   This is plotted in Fig.~\ref{lamb} for a state-of-the-art GaAs ($\epsilon=13$) PC cavity with $\omega_c/2\pi=200$\,THz, $Q=\omega_c/\Gamma_c=6000$, and $V_{\rm eff} = 0.063\,\mu m^3$ , containing a 30 Debye ($0.626\,{\rm e\,nm}$) QD \cite{Yao2009} at its antinode.  We note that the Lamb shift is symmetric, goes to zero on resonance, and has a peak amplitude which is proportional to $Q$ and inversely proportional to $V_{\rm eff}$.

In slow-light waveguide structures  the asymmetry of the resonances results in a  rich frequency dependence of the lamb shift \cite{Yao2009meta, Wang2004}, and it is interesting to explore such effects with our PC NW waveguides.  The Lamb shift experienced by a 30-debye QD is shown in Fig.~\ref{lamb} for waveguide designs of Secs.~\ref{subsec:idealfinite} and \ref{subsec:real}.  We note that in both cases the multiple resonances in the LDOS lead to a similarly multiply-peaked Lamb shift, and the overall asymmetry also produces a large dc competent.   The amplitude of the peaks is substantially lower in waveguides than the cavity example due to their $Q/V_{\rm eff}$ which is orders of magnitude lower, but the bandwidth of the effects much more rich if one properly accounts for the multi-modal nature of the photonic band structure.  The Lamb shifts at the primary resonance $\omega_{0}$ are calculated as $2.1$\,GHz and $4.95$\,THz for the idealized and standard structures, respectively (cf.~the simple cavity, which analytically 0\,GHz).  The former is comparable with the largest values reported in metameterial waveguides \cite{Yao2009meta} and PC structures \cite{Wang2004}; while  the DC component of the latter is orders of magnitude larger than previous reports \cite{Yao2009meta}, originating largely from the inclusion of the substrate.  Investigation of various other NW PC waveguides has indicated that the substrate introduces a rich modal structure far from the waveguide band resonances, all of which contribute to this large DC offset in the Lamb shift.  Furthermore, the QD location on top of the substrate was seen to increase coupling with odd modes, again resulting in a larger Lamb shift due to the large local field enhancements near the top of the NW, an effect unique to this platform.  These NW PC waveguides thus produce a rich and complex frequency and positional dependent Lamb shift, which can be exploited in the design of devices or measured as a test bed for waveguide QED.

\subsection{Photon Gun}
In this final subsection, we describe the design of a directed single photon source based on NW PC waveguides.  Up to this point, the $\beta$ factors given have been the probability of a single photon emitted from the QD exiting the structure via the waveguide mode, \textit{in either direction}.  In order to emit photons in a single direction, we truncate the NW waveguide in one direction with bulk PC NWs to form a photon gun, as was proposed for slab PC waveguides in Ref.~\onlinecite{Rao2007}.  If the emitter location is chosen carefully, constructive interference from reflections off the truncated waveguide-PC interface will increase the field strength, effectively doubling the Purcell factor.  It was found using FDTD simulations that an emitter in the central NW of the waveguide channel optimized this constructive interference.

\begin{figure}[h!]

\vspace{-5pt}
\subfloat[\vspace{-2pt}]{\label{pgunpfb}\includegraphics[width=0.49\textwidth]{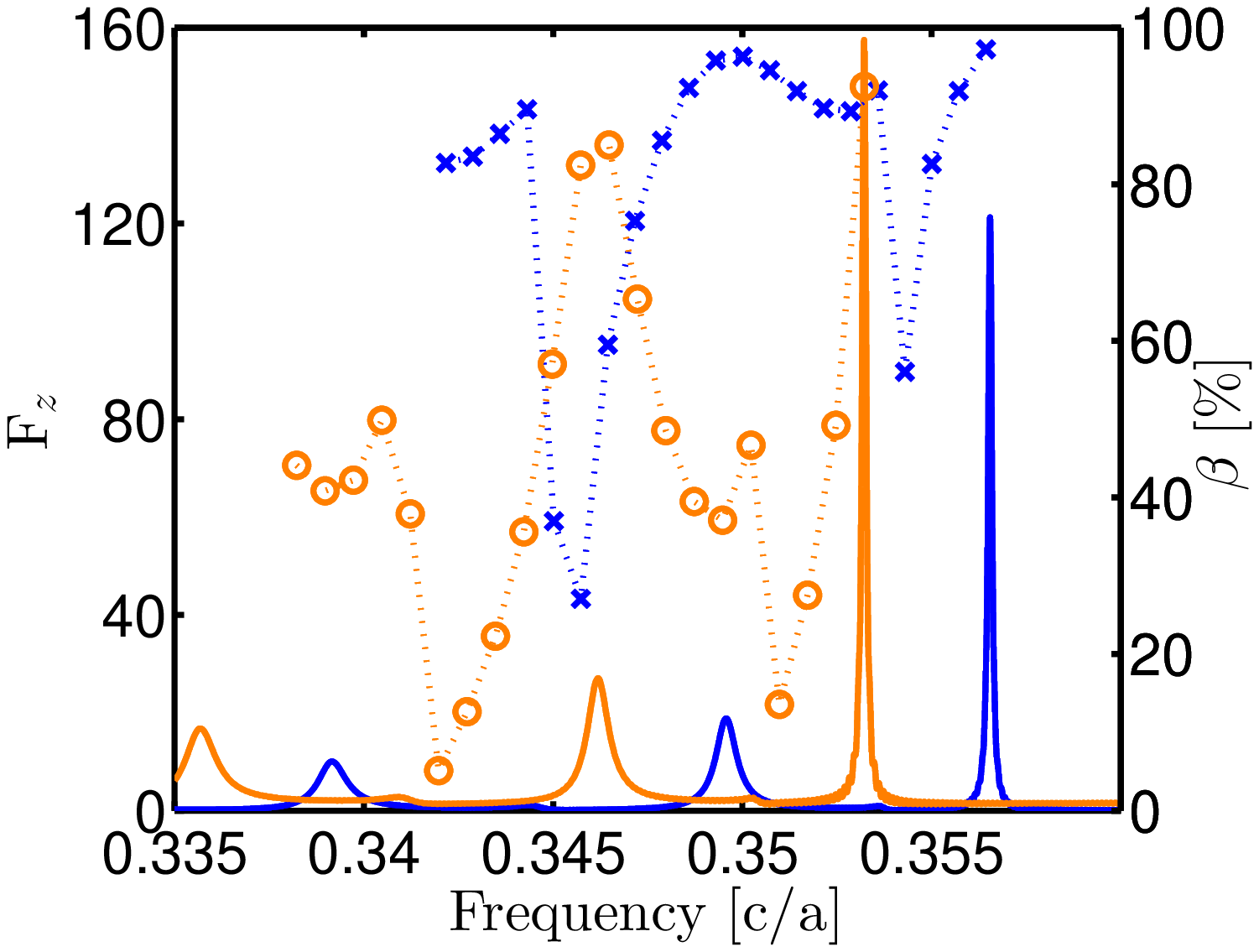}}

\vspace{-5pt}
\subfloat[$f\!=\! 0.3564\frac{c}{a}$\vspace{-2pt}]{\label{gunout1}\includegraphics[width=0.49\columnwidth]{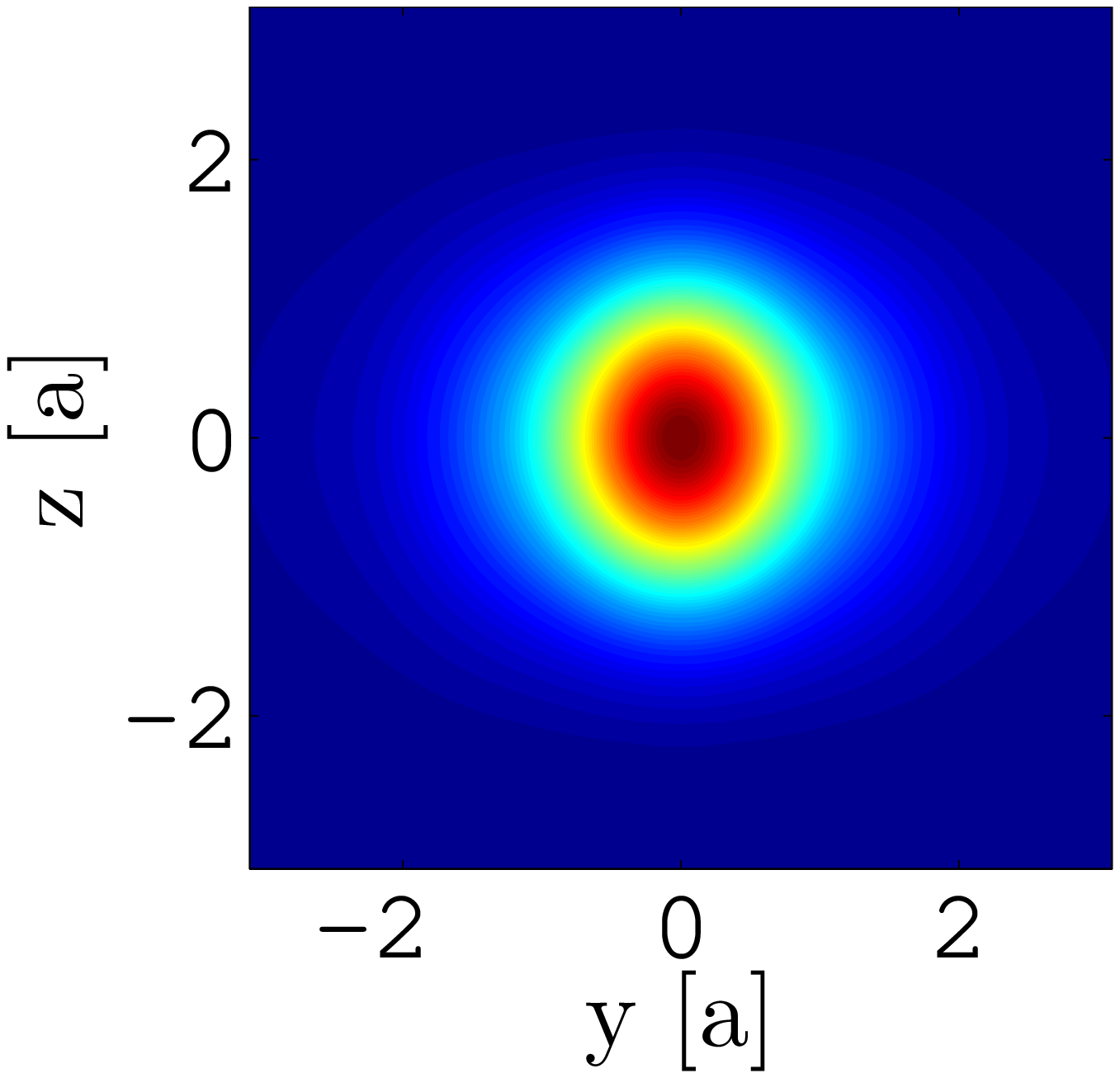}}
\subfloat[$f\!=\! 0.3532\frac{c}{a}$\vspace{-2pt}]{\label{gunout2}\includegraphics[width=0.49\columnwidth]{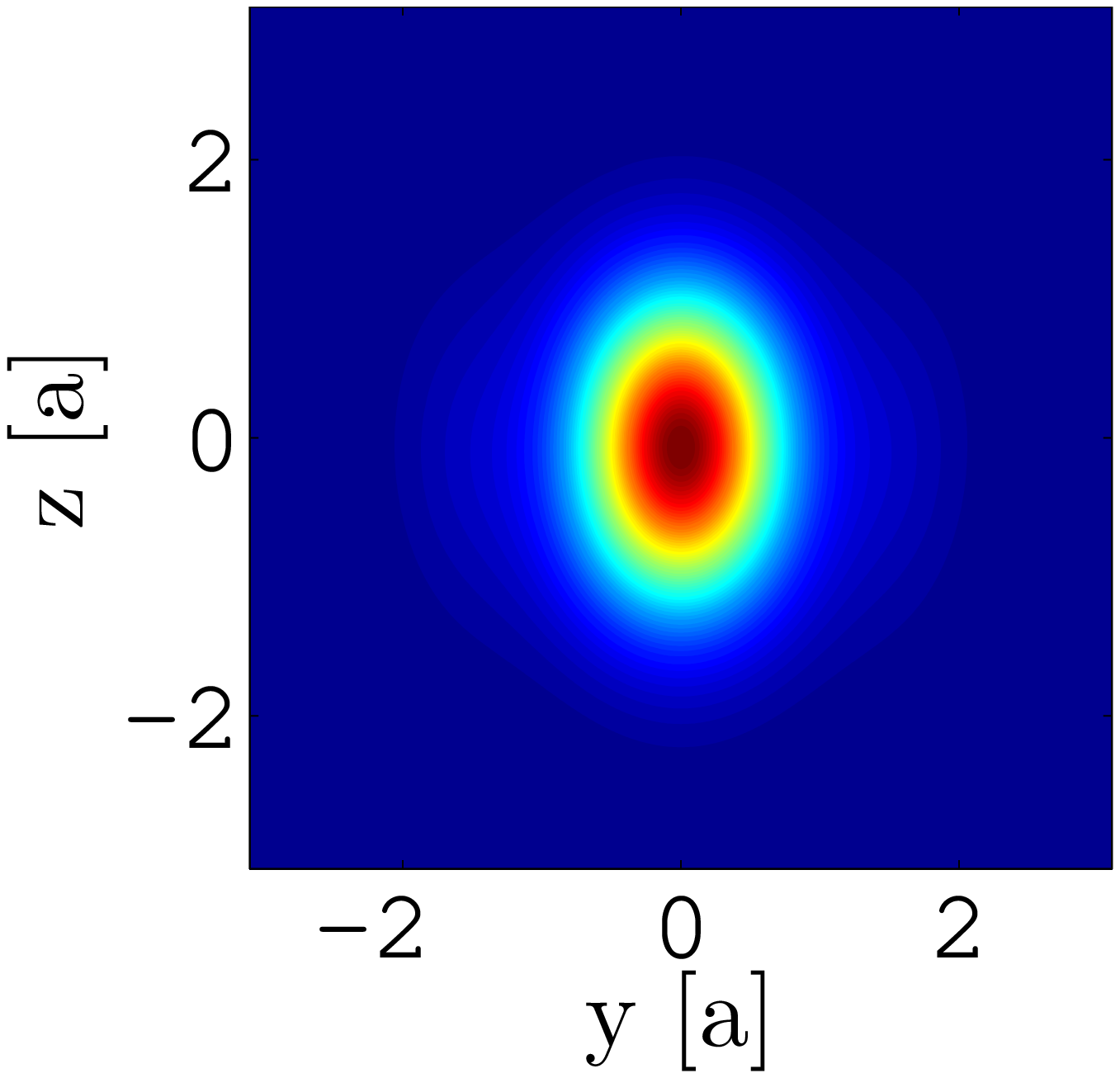}}

\vspace{-5pt}
\subfloat[$f\!=\!0.3532\frac{c}{a}$\vspace{-2pt}]{\label{topgun}\includegraphics[width=0.49\textwidth]{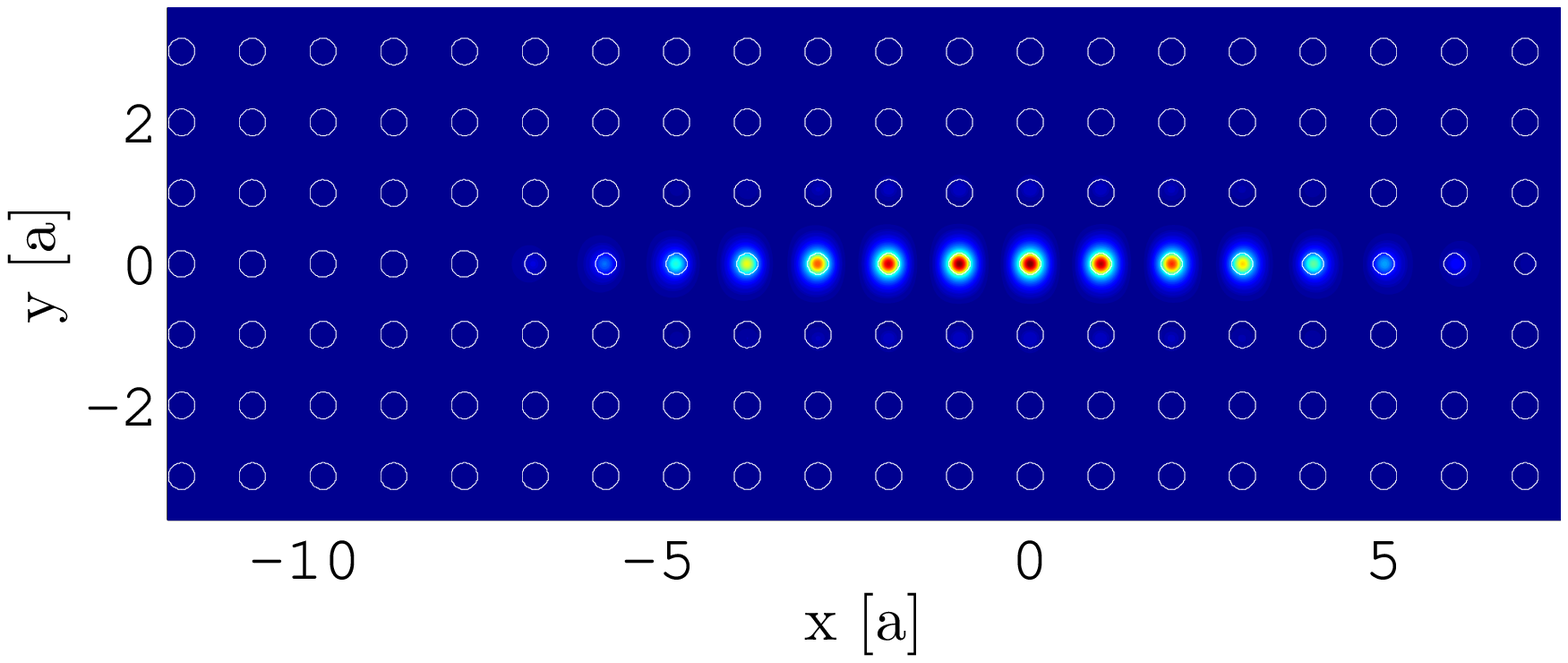}}

\vspace{-10pt}
\caption{(Color online) (a) Purcell and $\beta$ factors of realistic and substrate-free  photon gun in light orange  and dark blue, respectively.  Solid lines denote $F_z$, markers correspond to calculated $\beta$ factor values.  (b) and (c) $S_x$ in arbitrary units calculated $1.4a$ from the terminus of substrate-free and standard photon gun respectively, corresponding to power flow out of the structure.  (d) $|\mathbf{E}_{\lambda_0}|^2$  in the $z=0$ plane of the proposed single photon gun.}
\label{pgun}
\vspace{-10pt}
\end{figure}Two NW photon guns were studied, an idealistic one with a QD in the center of a NW and no substrate, and the more realistic system proposed in Sec.~\ref{subsec:real}.  Both structures contain a $15a$-length waveguide with the emitter in the central NW and truncated in one end with 5$a$ of bulk PC NWs, bringing the total length of the structure to 20$\,a$.  Their single photon properties are presented in Fig.~\ref{pgun}\subref{pgunpfb}, with the mode edge $Q$, $F_d$ and $f_{0}$ of the photon gun being 2730, 157.5, and 0.3532$\,\frac{c}{a}$ (2995, 121.1, and  0.3565$\,\frac{c}{a}$ for the ideal structure). As predicted \cite{Rao2007}, the PF more than doubles relative to the equivalent NW PC, and the mode edge also blueshifts slightly.  The calculated $\beta$ factors show far greater spread than previous structures, with the $\beta$ of the more realistic structure falling as low as 13.6\% before increasing to its peak of 92.5\% at the mode edge.  The low $\beta$ factors at select frequency points are likely due to destructive interference preventing certain modes from exiting the structure via the waveguide channel.  We note that the ideal substrate-free structure contains a broad range of $\beta>90\%$, and a peak value of 97.2\% at the mode edge, as emitted photons from a embedded QD are more likely to couple into the structure even if a strong waveguide resonance is not present.

 Finally, Figs.~\ref{pgun}\subref{gunout1} and  \ref{pgun}\subref{gunout2} show the power flow out of the ideal and elevated device, respectively, as measured 1.4$a$ ($\sim0.76\,{\rm \mu m}$) from the terminus of the photon gun structure.  A strongly localized profile is clearly visible in both cases, which can be readily collected by a detector or coupled into further optical components such as a conventional dielectric waveguide.  Figure~\ref{pgun}\subref{topgun} shows a vertical profile of the band edge quasi-mode of the realistic NW waveguide.  The waveguide mode is clearly reflected by the bulk NW section.

\section{Conclusions}
We have introduced and analyzed a new on-chip platform for studying open-system QED on a PC waveguide configuration that uses NW arrays with embedded QDs.   These NW PC systems produced waveguides with near unity $\beta$ factors over broadband frequencies and yield an enhanced emission factor exceeding 100 even in small realistic devices; we also proposed a  photon gun with single photon source parameters exceeding those in the best  slab PCs \cite{Rao2007}.  In addition, we showed that interesting and measurable Lamb shifts are produced in these NW PC structures.  This nanowire PC platform has the potential to implement more complex integrated systems for studying and exploiting quantum optical effects, e.g., using multiple QDs coupled on the same waveguide.  
\vspace{5pt}
\begin{acknowledgements}
 We thank Ray LaPierre and Henry Schriemer for useful discussions, and  Nishan Mann and  Rong-Chun Ge for assistance with computational software.  This work was supported by the Natural Science and Engineering Research Council of Canada. 
\end{acknowledgements}
\bibliography{paperbib2}
\end{document}